\documentclass[referee]{raa}
\usepackage{graphicx,times}
\usepackage{natbib}
\usepackage{amssymb,amsmath}
\usepackage{longtable}
\bibpunct{(}{)}{;}{a}{}{,}

\usepackage[a4paper=true,dvipdfm=true,pagebackref=true]{hyperref}
\hypersetup{pdftitle = The title of my PDF, pdfauthor = My name, pdfsubject= The subject, pdfkeywords = keyword1 keyword2 keyword3}
\hypersetup{colorlinks = true, linkcolor = green, anchorcolor = red, citecolor = blue, filecolor = red, pagecolor = red, urlcolor = red}

\begin{document}

   \title{Light curve solutions of six eclipsing binaries at the lower limit of periods of the W UMa stars}

 \volnopage{ {\bf 2012} Vol.\ {\bf X} No. {\bf XX}, 000--000}
   \setcounter{page}{1}

   \author{Diana P. Kjurkchieva\inst{1}, Dinko P. Dimitrov\inst{2} and Sunay I. Ibryamov\inst{1,2}}

   \institute{Department of Physics, Shumen University, 115, Universitetska Str., 1712 Shumen, Bulgaria; {\it d.kyurkchieva@shu-bg.net}\\
                \and
              Institute of Astronomy and National Astronomical Observatory, Bulgarian Academy of Sciences, 72, Tsarigradsko Shose Blvd., 1784 Sofia, Bulgaria\\
   {\small Received ; accepted }}
	
	\abstract{Photometric observations in \emph{V} and \emph{I} bands of six eclipsing binaries at the lower limit of the orbital periods of W UMa stars are presented. Three of them are newly discovered eclipsing systems. The light curve solutions revealed that all short-period targets were contact or overcontact binaries and added new six binaries to the family of short-period systems with estimated parameters. Four binaries have equal in size components and mass ratio near 1. The phase variability of the \emph{V-I} colors of all targets may be explained by lower temperatures of their back surfaces than those of their side surfaces. Five systems revealed O'Connell effect that was reproduced by cool spots on the side surfaces of their primary components. The light curves of V1067 Her in 2011 and 2012 were fitted by diametrically opposite spots. The applying of the criteria for subdivision of the W UMa stars to our targets led to ambiguous results.
 }

   \authorrunning{Kjurkchieva, Dimitrov, Ibryamov}            
   \titlerunning{Light curve solutions of six eclipsing binaries at the low period limit of the W UMa stars}  
   \maketitle

%
\section{Introduction}           
\label{sect:intro}

Most of the W UMa stars consisting of solar-type components have orbital periods within 0.25 d $< P <$ 0.7 d. They are recognized by continuous brightness variations and nearly equal minima. The short orbital periods of these binaries mean small orbits and synchronized rotation and orbital revolution.


\newpage
The statistics of W UMa stars around the lower limit of periods is quite poor (\citealt{Terrell+etal+2012}) because the period distribution of binaries reveals a very sharp decline below 0.27 days (\citealt{Drake+etal+2014}). Another reason is that the short-period binaries consist of late stars and thus are faint objects for detailed study.

However, the short-period contact systems are important objects for the modern astrophysics at least in two ways: (a) the empirical relation period-luminosity of W UMa's allows to use them as distance and population tracers; (b) these binaries are natural laboratories for study of late stage of the stellar evolution connected with the processes of mass and angular momentum loss, merging or fusion of the stars, etc.

Lately, the interest to binaries around the low-period limit increases because they provide constraints on the theories for the formation and evolution of the low-mass stars. Moreover, one of the hypotheses for the origin of the hot Jupiters is that they are products of proto-planetary disk formed by merging of short-period, low-mass binaries via ''magnetic braking'' if substantial amount of mass remains in orbit around the primary (\citealt{Martin+etal+2011}).

Fortunately, the modern large stellar surveys during the last decade allowed to discover binaries with shorter and shorter periods \cite{Rucinski+2007}, \cite{Pribulla+etal+2009}, \cite{Weldrake+etal+2004}, \cite{Maceroni+Rucinski+1997}, \cite{Dimitrov+Kjurkchieva+2010}, \cite{Norton+etal+2011}, \cite{Nefs+etal+2012}, \cite{Davenport+etal+2013}, \cite{Lohr+etal+2014}, \cite{Qian+etal+2014}, \cite{Drake+etal+2014}. The majority of the newly discovered binaries from space missions (\emph{Kepler, CoRoT}, etc.) and ground-based projects (ASAS, SuperWASP, Catalina, LINEAR, NSVS, etc.) were classified as contact systems. But most of them need follow-up observations and study.

In this paper we present photometric observations and light curve solutions of six binaries around the lower limit of the orbital periods of W UMa stars. We chose to study such systems because the statistics of W UMa stars with periods around quarter day is quite poor (\citealt{Terrell+etal+2012}). Three of our targets are known binaries: two binaries (1SWASP J173828.46+111150.2, 1SWASP J174310.98+432709.6 $\equiv$ V1067 Her) are from the SuperWASP photometric survey (\citealt{Pollacco+etal+2006}) and one system (NSVS 11534299) is from the NSVS data base (\citealt{Wozniak+etal+2004}). Three of our targets (NSVS 11234970, NSVS 11504202, NSVS 10971359) are newly discovered binaries. Table~\ref{Tab1} presents the coordinates of our targets and information for their light variability.

\begin{table}
\bc
\begin{minipage}[]{100mm}
\caption[]{List of our targets\label{Tab1}}\end{minipage}
\setlength{\tabcolsep}{1pt}
\small
 \begin{tabular}{cccccccc}
  \hline\noalign{\smallskip}
Target & Name & $\alpha_{2000}$ & $\delta_{2000}$ & Period [d] & $V$ [mag] & $\Delta V$ & Ref\\
  \hline\noalign{\smallskip}
1 & 1SWASP J173828.46+111150.2 & 17 38 28.459 & +11 11 50.05 & 0.249383 & 13.80& 0.24 & 1\\

2 & 1SWASP J174310.98+432709.6 & 17 43 10.977 & +43 27 09.48 & 0.258108 & 13.20& 0.44 & 2\\

3 & NSVS 11234970              & 19 32 07.875 & +14 58 26.90 & 0.25074  & 13.66& 0.63 & 3\\

4 & NSVS 11504202              & 20 27 27.304 & +12 30 50.66 & 0.24614  & 13.67& 0.66 & 3\\

5 & NSVS 11534299              & 20 48 20.859 & +06 59 22.98 & 0.22465  & 13.40& 0.21 & 4\\

6 & NSVS 10971359              & 18 29 20.479 & +07 58 07.26 & 0.27977  & 13.31& 0.38 & 3\\
  \noalign{\smallskip}\hline
\end{tabular}
\ec

References: 1 -- \cite{Lohr+etal+2012}; 2 -- \cite{Blattler+Diethelm+2000}; 3 -- this paper; 4 -- \cite{Hoffman+etal+2009}
\end{table}

\newpage
\section{Observations}

The CCD photometry of the targets was carried out in 2011$-$2012 at Rozhen National Astronomical Observatory (Bulgaria) with the 60-cm Cassegrain telescope using the FLI PL09000 CCD camera (3056 $\times$ 3056 pixels, 12 $\mu m/$pixel, field of 17.1\arcmin $\times$ 17.1\arcmin). The average photometric precision per data point is 0.005 mag in \emph{I} band and 0.008 mag in \emph{V} band. Table~\ref{Tab2} presents the journal of our simultaneous $VI$ observations. In addition we obtained several observations in $BVI$ colors of the targets 1, 3 and 4.

\begin{figure}
   \centering
   \includegraphics[width=13.0cm, angle=0]{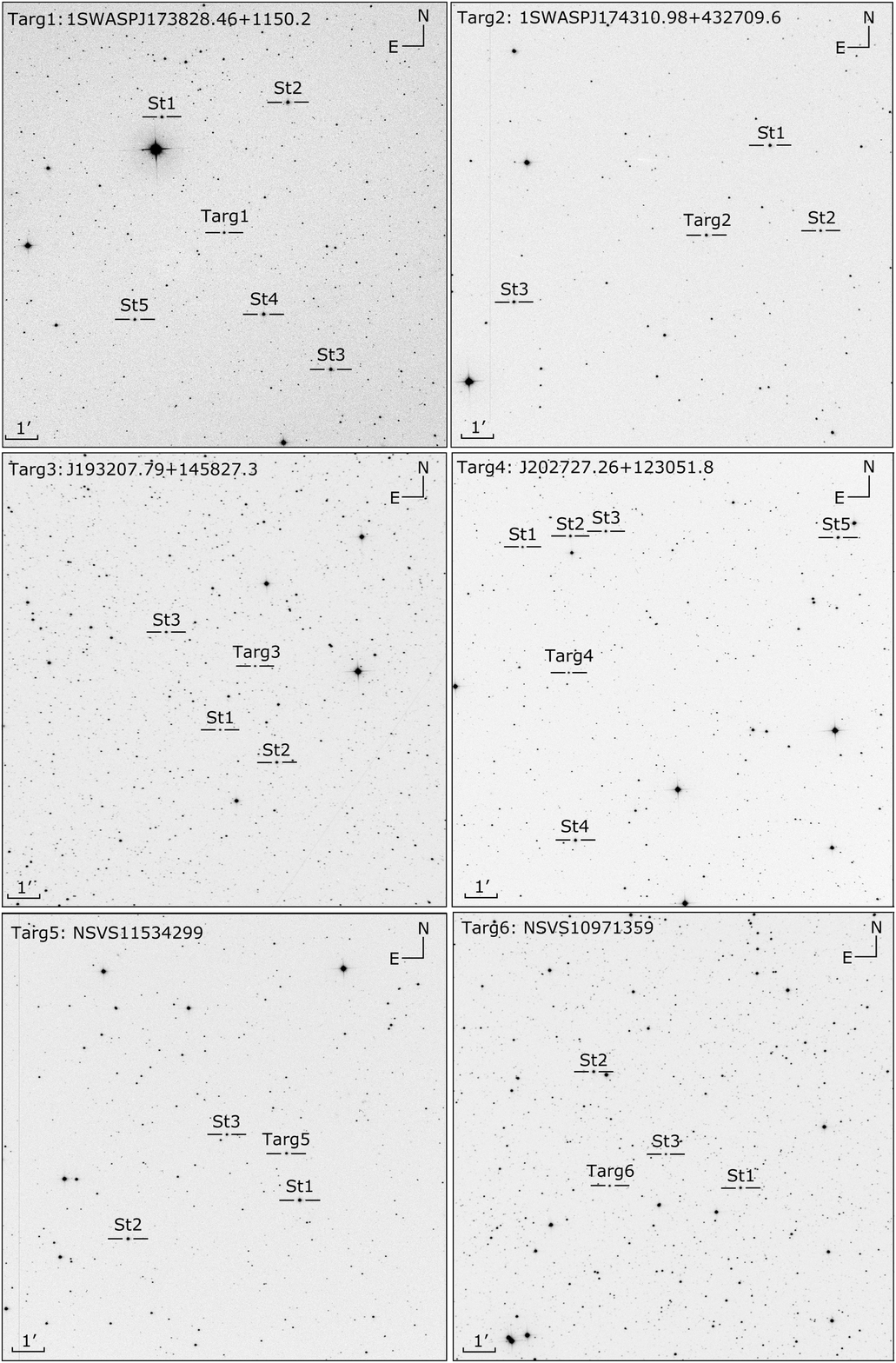}
   \caption{The fields of the targets}
   \label{Fig1}
   \end{figure}

The photometric data were reduced by IDL package (subroutine DAOPHOT). We used more than three standard stars in the observed fields (Fig.~\ref{Fig1}). Table~\ref{Tab3} presents their colors: \emph{I} from USNO-B1.0 catalogue and \emph{V} from GSC 2.3.2 catalogue. The magnitudes of the targets in Table~\ref{Tab3} correspond to their out-of-eclipse levels from our observations.

\begin{table}
\bc
\begin{minipage}[]{100mm}
\caption[]{Journal of the Rozhen photometric observations\label{Tab2}}\end{minipage}
\setlength{\tabcolsep}{1pt}
\small
 \begin{tabular}{ccccc}
  \hline\noalign{\smallskip}
Target&  Date  & Filter & Exposure [sec] & Phase range\\
  \hline\noalign{\smallskip}
1 & 2011 July 10 & $V, I$ & 120,120 & 0.21-1.17 \\
  & 2011 Aug 03  & $V, I$ & 120,120 & 0.67-0.89 \\
  & 2011 Aug 21  & $V, I$ & 120,120 & 0.60-0.94 \\
  & 2012 June 02 & $V, I$ & 120,120 & 0.61-1.41 \\
  & 2012 June 03 & $V, I$ & 120,120 & 0.24-1.35 \\
  & 2012 July 07 & $V, I$ & 120,120 & 0.90-1.74 \\
2 & 2011 Aug 11  & $V, I$ & 120,120 & 0.20-1.08 \\
  & 2011 Aug 19  & $V, I$ & 120,120 & 0.83-1.74 \\
  & 2012 June 04 & $V, I$ & 120,120 & 0.86-1.33 \\
  & 2012 July 08 & $V, I$ & 120,120 & 0.35-1.38 \\
3 & 2012 Aug 04  & $V, I$ &  90,120 & 0.60-1.94 \\
4 & 2012 Aug 05  & $V, I$ & 120,120 & 0.63-1.91 \\
5 & 2012 Aug 06  & $V, I$ & 120,120 & 0.19-1.32 \\
6 & 2011 July 08 & $V, I$ & 120,120 & 0.45-1.24 \\
  & 2011 July 11 & $V, I$ & 120,120 & 0.16-0.90 \\
  \noalign{\smallskip}\hline
\end{tabular}
\ec
\end{table}

We carried out preliminary time-series analysis of our data by the software \emph{PerSea}. The comparison of the results with the published ones (Table~\ref{Tab1}) revealed two discrepancies.

(a) The periods of targets 1 and 2 phased well our data. But the period 0.24914 d of target 5 (see further Table~\ref{Tab5}) we obtained was around 11 $\%$ longer than the previous value (Table~\ref{Tab1}). This result is not so surprising because the follow-up observations of variable stars from different data bases sometimes reveal different periods than the previous values (\citealt{Norton+etal+2011}; \citealt{Mighell+Plavchan+2013}; \citealt{Lohr+etal+2013}; \citealt{Kjurkchieva+etal+2014}; \citealt{Mayangsari+etal+2014}). This is a consequence of the automated frequency analyses of the huge data sets. They based on different approaches which turned out with different precision for different types of variabilities. Moreover, sometimes they give several periods for each target (as well as false alarm detections of exoplanets).

\newpage

(b) The amplitudes of variability of targets 1 and 2 turned out considerably bigger than the previous ones. We assume that the reason for this discrepancy is the low space resolution of the previous photometric observations by small telescopes (13.7\arcsec pixel$^{-1}$ for SWASP survey and 14.4\arcsec pixel$^{-1}$ for NSVS) that prevents separation of two neighboring stars.

\newpage
The close nonvariable star may enter the same pixel or aperture during the photometric measurements of the variable and thus to reduce its amplitude of variability.

\begin{table}
\bc
\begin{minipage}[]{100mm}
\caption[]{Magnitudes of the targets and their standard stars\label{Tab3}}\end{minipage}
\setlength{\tabcolsep}{1pt}
\small
 \begin{tabular}{llccc}
  \hline\noalign{\smallskip}
Star         & GSC-ID       & $I$   & $V$   & $B$ \\
             & (USNO B1)    & [mag] & [mag] &[mag]\\
  \hline\noalign{\smallskip}
targ         & 1            & 12.21 & 13.46 & 14.68\\
st1          & 0100100125   & 11.55 & 12.67 & 13.41\\
st2          & 0100101007   & 10.83 & 12.20 & 13.62\\
st3          & 0099702419   & 11.04 & 12.49 & 14.15\\
st4          & 0099702343   & 11.10 & 12.57 & 14.09\\
st5          & 0099702387   & 11.85 & 13.12 & 14.10\\
targ         & 2            & 12.04 & 12.82 &      \\
st1          & 0310001679   & 12.49 & 13.00 &      \\
st2          & 0310001797   & 13.45 & 13.66 &      \\
st3          & 0310001604   & 11.13 & 12.01 &      \\
targ         & 3            & 12.45 & 13.66 & 14.33\\
st1          & 1049-0454686 & 12.88 & 13.44 & 13.87\\
st2          & 1049-0454529 & 10.91 & 12.45 & 13.33\\
st3          & 1049-0454842 & 11.75 & 12.88 & 13.41\\
targ         & 4            & 12.43 & 13.67 & 14.64\\
st1          & 0109500807   & 12.63 & 13.19 & 14.79\\
st2          & 0109501223   & 10.98 & 12.27 & 13.70\\
st3          & 0109500601   & 11.22 & 12.78 & 14.65\\
st4          & 0109501589   & 10.66 & 11.33 & 11.93\\
st5          & 0109500781   & 10.88 & 12.38 & 14.02\\
targ         & 5            & 11.97 & 13.19 &      \\
st1          & 0052402305   & 10.71 & 12.13 &      \\
st2          & 0052401323   & 11.50 & 11.93 &      \\
st3          & 0052401783   & 11.94 & 13.32 &      \\
targ         & 6            & 12.92 & 13.31 &      \\
st1          & 0102300733   & 12.14 & 12.34 &      \\
st2          & 0102302368   & 10.87 & 11.93 &      \\
st3          & 0979-0436284 & 13.77 & 13.97 &      \\
  \noalign{\smallskip}\hline
\end{tabular}
\ec
\end{table}

\section{Light curve solutions}

\begin{figure}
   \centering
   \includegraphics[width=7.5cm, angle=0]{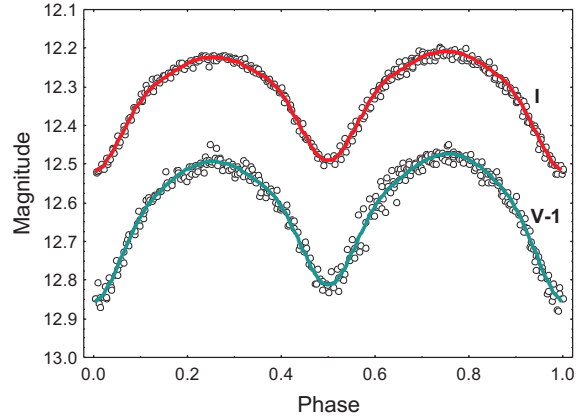}
   \caption{Light curves of target 1 in $V$ and $I$ bands (the \emph{V} data are shifted vertically by 1$^{m}$ for better visibility) and their fits}
   \label{Fig2}
   \end{figure}

\begin{figure}
   \centering
   \includegraphics[width=7.5cm, angle=0]{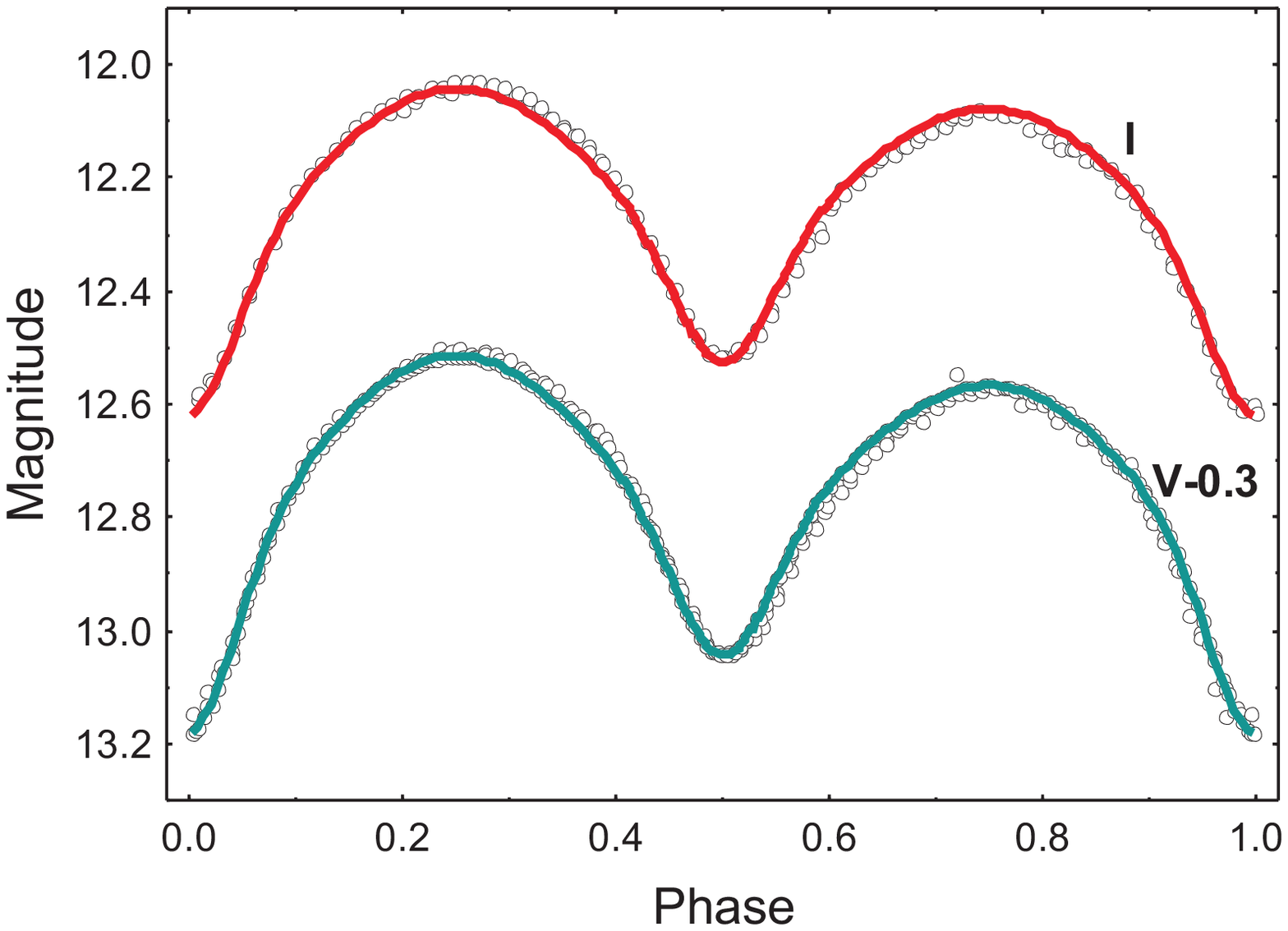}
   \caption{Light curves of target 2 in $V$ and $I$ bands in 2011 (the \emph{V} data are shifted vertically by 0.3$^{m}$ for better visibility) and their fits}
   \label{Fig3}
   \end{figure}

\begin{figure}
   \centering
   \includegraphics[width=7.5cm, angle=0]{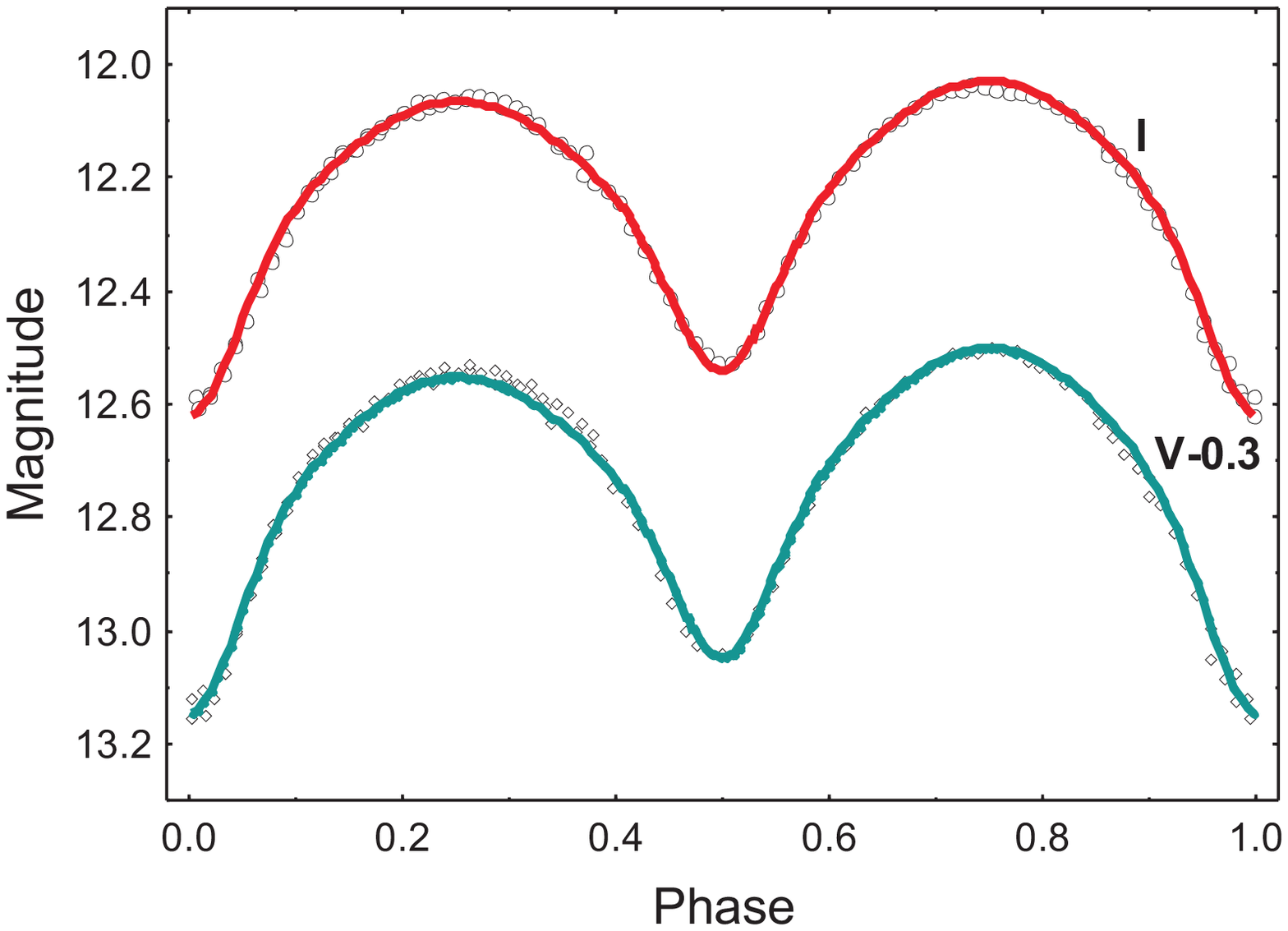}
   \caption{Light curves of target 2 in $V$ and $I$ bands in 2012 (the \emph{V} data are shifted vertically by 0.3$^{m}$ for better visibility) and their fits}
   \label{Fig4}
   \end{figure}

\begin{figure}
   \centering
   \includegraphics[width=7.5cm, angle=0]{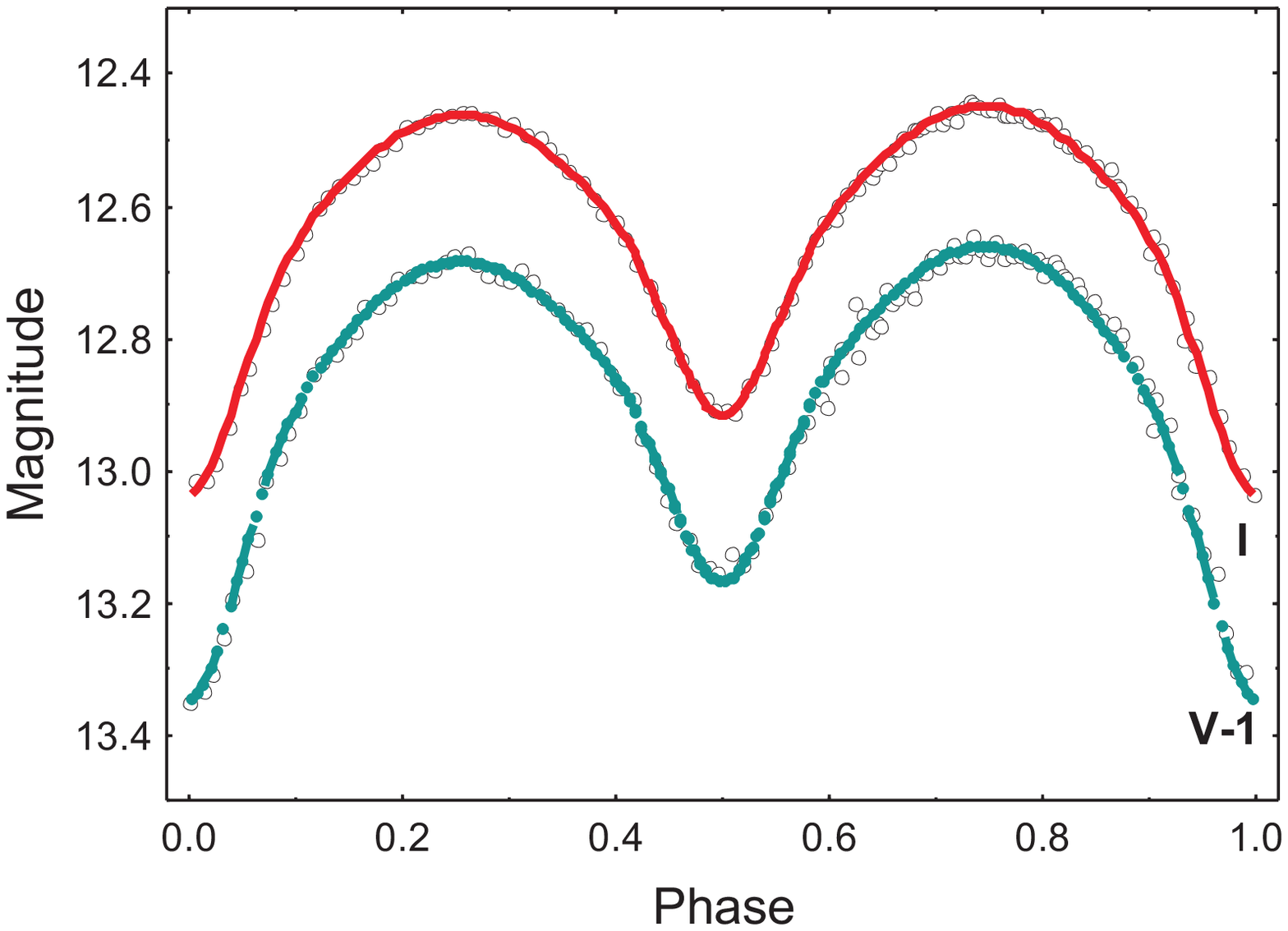}
   \caption{Light curves of target 3 in $V$ and $I$ bands (the \emph{V} data are shifted vertically by 1$^{m}$ for better visibility) and their fits}
   \label{Fig5}
   \end{figure}

\begin{figure}
   \centering
   \includegraphics[width=7.5cm, angle=0]{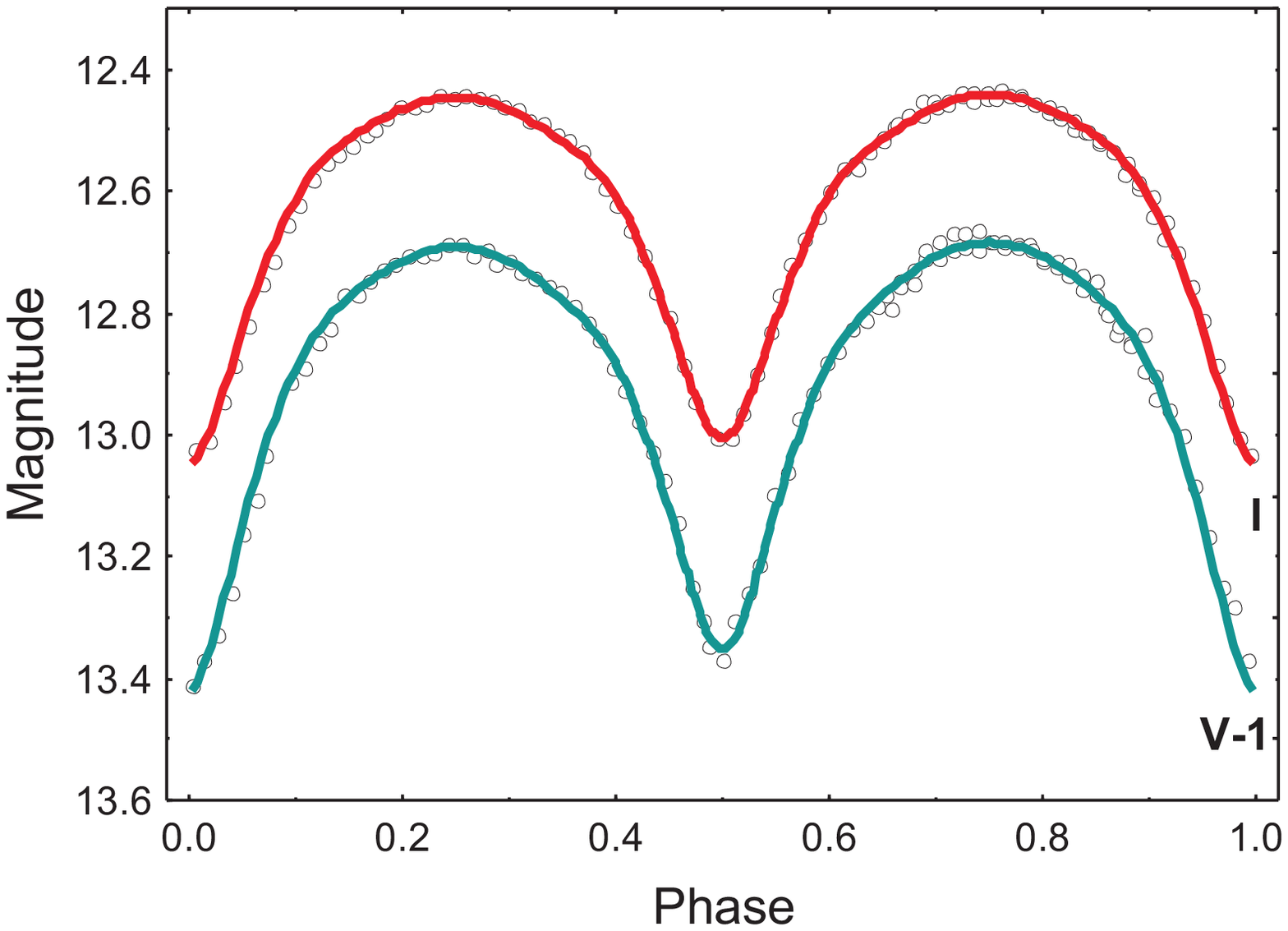}
   \caption{Light curves of target 4 in $V$ and $I$ bands (the \emph{V} data are shifted vertically by 1$^{m}$ for better visibility) and their fits}
   \label{Fig6}
   \end{figure}

\begin{figure}
   \centering
   \includegraphics[width=7.5cm, angle=0]{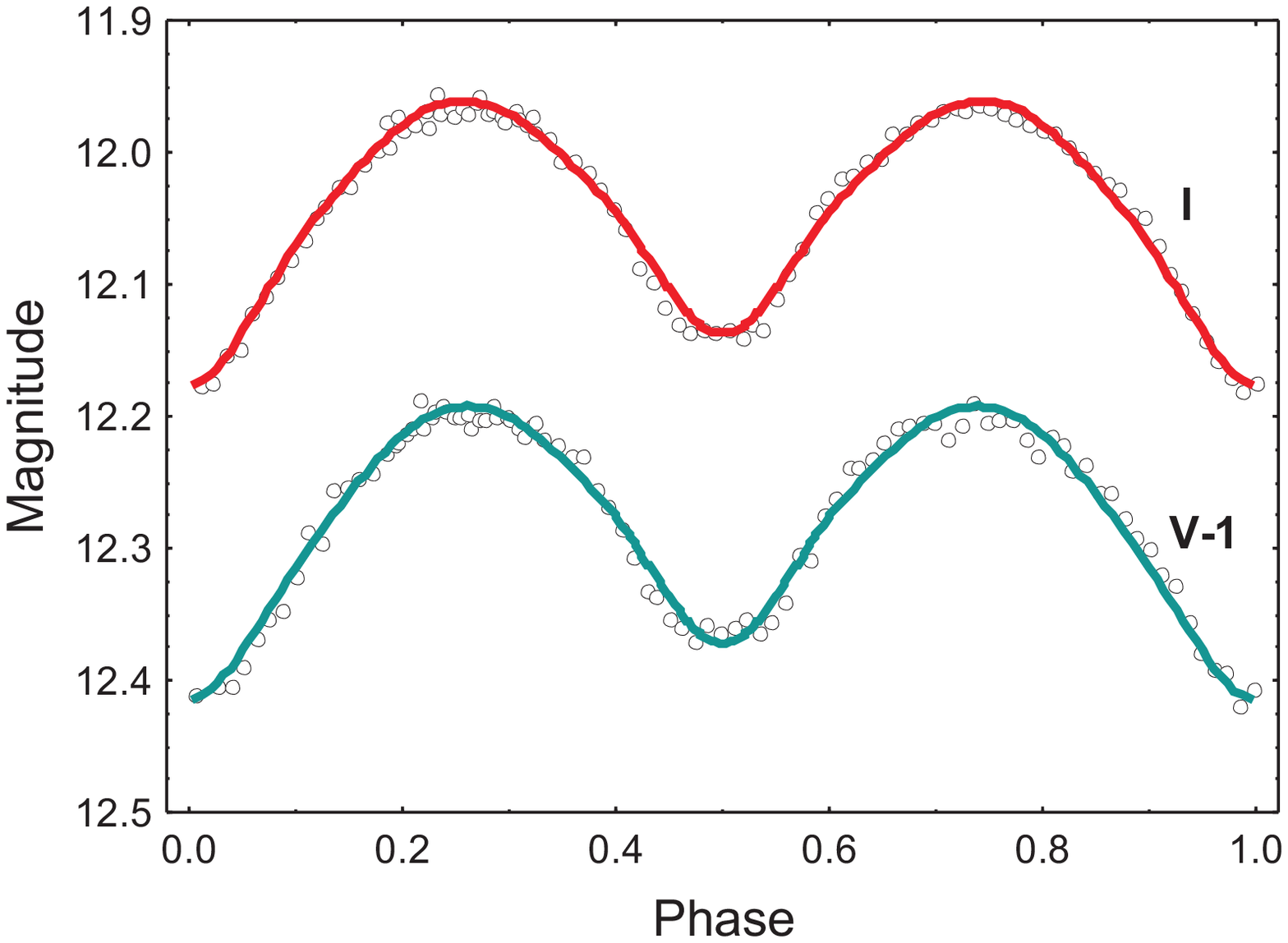}
   \caption{Light curves of target 5 in $V$ and $I$ bands (the \emph{V} data are shifted vertically by 1$^{m}$ for better visibility) and their fits}
   \label{Fig7}
   \end{figure}

The shapes of the light curves (Figs.~\ref{Fig2},~\ref{Fig3},~\ref{Fig4},~\ref{Fig5},~\ref{Fig6},~\ref{Fig7}) implied that our targets are nearly contact or overcontact systems that was expected for their short orbital periods. The big amplitudes of their light variabilities mean that they are caused by eclipses.

It is well known that the determination of the mass ratio through light-curve solution is an ambiguous approach compared with that by radial velocity solution. 

\newpage
However, the rapid rotation of the components of the short-period binaries is serious obstacle to obtain precise spectral mass ratio from measurement of their
highly broadened and blended spectral lines (\citealt{Bilir+etal+2005}, \citealt{Dall+Schmidtobreick+2005}). On the other hand, their eclipse depths depend strongly on the potentials and the mass ratios.

Taking into account these considerations we solved the Rozhen light curves of the six targets using the code \emph{PHOEBE} (\citealt{Prsa+Zwitter+2005}) by the following procedure.

\begin{table}
\bc
\begin{minipage}[]{100mm}
\caption[]{The 2 MASS color indices \emph{J-K} and corresponding mean temperatures T$_{m}$ of the targets\label{Tab4}}\end{minipage}
\setlength{\tabcolsep}{1pt}
\small
 \begin{tabular}{lccccccc}
  \hline\noalign{\smallskip}
Target  & 1               & 2               & 3               & 4               & 5               & 6  \\
  \hline\noalign{\smallskip}
\emph{J-K  }   & 0.645$\pm$0.033 & 0.590$\pm$0.024 & 0.749$\pm$0.030 & 0.750$\pm$0.026 & 0.558$\pm$0.034 & 0.618$\pm$0.031 \\
T$_{m}$ & 4720$\pm$120    & 4930$\pm$140    & 4320$\pm$130    & 4320$\pm$110    & 5090$\pm$220    & 4810$\pm$130  \\
  \noalign{\smallskip}\hline
\end{tabular}
\ec
\end{table}
%

We determined in advance the mean temperatures $T_{m}$ of the binaries (Table~\ref{Tab4}) by their infrared color indices \emph{(J-K)} from the 2MASS catalog and the calibration color-temperature of \cite{Tokunaga+2000}.

At the first stage we adopted $T_{1}$=$T_{m}$ (a good approximation for contact binaries with close temperatures of the components) and searched for solutions for fixed $T_{1}$ varying the initial epoch $T_{0}$, period $P$, secondary temperature $T_{2}$, orbital inclination $i$, mass ratio $q$ and potentials
$\Omega_{1,2}$.

We adopted coefficients of gravity brightening $g_1=g_2$ = 0.32 and reflection effect $A_1=A_2$ = 0.5 appropriate for late stars while the limb-darkening coefficients for each component and each color were auto-updated to the stellar temperature according to the tables of \cite{VanHamme+1993}.

In order to reproduce the O'Connell effect we added cool spots on the stellar surfaces and varied spot parameters (longitude $\lambda$, latitude $\beta$, angular size $\alpha$ and temperature factor $\kappa$). Moreover, the best fit in the two colors required a small contribution of third light $l_{3}$.

Finally, we searched for the best fits for fixed $q$, spot parameters and third light contributions by adjusting the primary temperature $T_{1}$ slightly above $T_{m}$ and changed correspondingly $T_{2}$, $\Omega_{1,2}$ and $i$.

\begin{figure}
   \centering
   \includegraphics[width=7.5cm, angle=0]{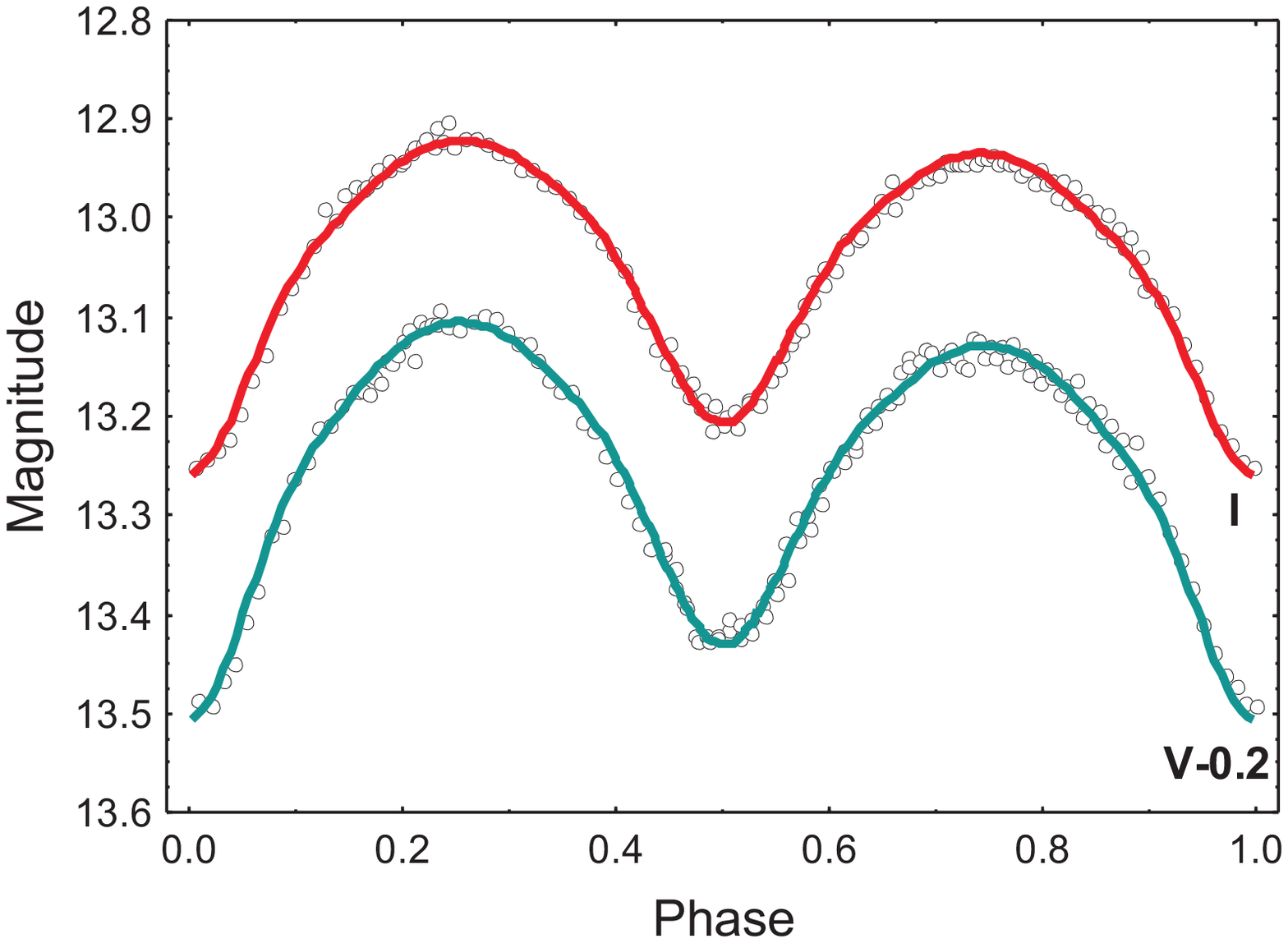}
   \caption{Light curves of target 6 in $V$ and $I$ bands (the \emph{V} data are shifted vertically by 0.2$^{m}$ for better visibility) and their fits}
   \label{Fig8}
   \end{figure}

Table~\ref{Tab5} contains the results of our light curve solutions. The parameter errors are the formal \emph{PHOEBE} errors. The synthetic light curves corresponding to the parameters from Table~\ref{Tab5} are shown in Figs.~\ref{Fig2},~\ref{Fig3},~\ref{Fig4},~\ref{Fig5},~\ref{Fig6},~\ref{Fig7},~\ref{Fig8}
as continuous lines.

    \begin{table}
\bc
\begin{minipage}[]{100mm}
\caption[]{Parameters of the light curve solutions of the targets\label{Tab5}}\end{minipage}
\setlength{\tabcolsep}{1pt}
\small
 \begin{tabular}{cllllll}
  \hline\noalign{\smallskip}
Parameter        & 1               & 2               & 3               & 4               & 5               & 6              \\
  \hline\noalign{\smallskip}
$T_{0}$ 2450000+ & 6091.94682      & 5785.294939     & 6144.364434     & 6145.36124      & 6146.46937      & 5751.49320    \\
                 & $\pm$0.00021    & $\pm$0.000151   & $\pm$0.000208   & $\pm$0.00011    & $\pm$0.00021    & $\pm$0.00021  \\
$   P   $        & 0.249383        & 0.258108        & 0.25074         & 0.24614         & 0.24914         & 0.27977        \\
                 &                 &                 & $\pm$0.00027    & $\pm$0.00024    & $\pm$0.00057    & $\pm$0.00041   \\
$   q   $        & 0.488$\pm$0.002 & 1.009$\pm$0.005 & 0.986$\pm$0.006 & 0.980$\pm$0.006 & 0.872$\pm$0.003 & 0.462$\pm$0.002\\
$   i   $        & 66.85$\pm$0.09  & 75.68$\pm$0.21  & 74.36$\pm$0.26  & 79.4$\pm$0.2    & 56.3$\pm$0.4    & 66.8$\pm$0.2   \\
$  T_1  $        & 4900$\pm$48     & 5152$\pm$94     & 4600$\pm$90     & 4402$\pm$55     & 5150$\pm$61     & 4936$\pm$48 \\
$  T_2  $        & 4544$\pm$28     & 4626$\pm$56     & 3969$\pm$70     & 4219$\pm$54     & 5014$\pm$45     & 4423$\pm$37    \\
$\Omega_1$       & 2.843$\pm$0.004 & 3.640$\pm$0.012 & 3.615$\pm$0.009 & 3.717$\pm$0.013 & 3.548$\pm$0.014 & 2.785$\pm$0.005  \\
$\Omega_2$       & 2.843$\pm$0.004 & 3.640$\pm$0.012 & 3.615$\pm$0.009 & 3.717$\pm$0.013 & 3.550$\pm$0.008 & 2.785$\pm$0.005  \\
$r_1$            & 0.445           & 0.400           & 0.399           & 0.363           & 0.389           & 0.451          \\
$r_2$            & 0.319           & 0.399           & 0.396           & 0.338           & 0.365           & 0.317         \\
$l_1$            & 0.746           & 0.643           & 0.715           & 0.598           & 0.566           & 0.787         \\
$l_2$            & 0.254           & 0.357           & 0.285           & 0.402           & 0.434           & 0.213         \\
$l_2$/$l_1$      & 0.340           & 0.555           & 0.399           & 0.672           & 0.767           & 0.271         \\
fillout          & 0.033           & 0.229           & 0.209           & -0.045          & -0.0024         & 0.061            \\
fillout          & 0.033           & 0.229           & 0.209           & -0.045          & -0.003          & 0.061            \\
configuration    & OC              & OC              & OC              & $\simeq$CB      & $\simeq$CB      & OC            \\
  \noalign{\smallskip}\hline
\end{tabular}
\ec
\end{table}

\begin{table}
\bc
\begin{minipage}[]{100mm}
\caption[]{Parameters of the cool spots on the targets\label{Tab6}}\end{minipage}
\setlength{\tabcolsep}{1pt}
\small
 \begin{tabular}{cccccc}
  \hline\noalign{\smallskip}
Target    & $\beta$   & $\lambda$ & $\alpha$  & $\kappa$      & l$_{3}$  \\
  \hline\noalign{\smallskip}
1         & 122$\pm$1 & 297$\pm$1 & 17$\pm$1  & 0.83$\pm$0.01 & 0.029 (I)\\
2$_{2011}$&  90$\pm$1 &  90$\pm$1 & 20$\pm$1  & 0.80$\pm$0.01 & 0.066 (I)\\
2$_{2012}$&  90$\pm$1 & 270$\pm$1 & 20$\pm$1  & 0.80$\pm$0.01 & 0.034 (I)\\
3         &  90$\pm$1 & 270$\pm$1 & 18$\pm$1  & 0.90$\pm$0.01 & 0.055 (I)\\
4         &  90$\pm1$ & 270$\pm$1 & 10$\pm$1  & 0.90$\pm$0.01 & 0.085 (I)\\
5         &  40$\pm$1 & 0$\pm$1   & 20$\pm$1  & 0.88$\pm$0.01 & 0.073 (V)\\
6         &  90$\pm$1 & 90$\pm$1  & 15$\pm$1  & 0.90$\pm$0.01 & 0.057 (I)\\
  \noalign{\smallskip}\hline
\end{tabular}
\ec
\end{table}

The results of our light curve solutions can be used for estimation of the global parameters of the targets (masses, radii and luminosities) using statistical relations and supposition that their components are approximate MS stars.

\section{Analysis of the results}

The analysis of the light curve solutions led to several conclusions.

\newpage
(1) All targets are almost contact ($\simeq$CB) or overcontact binaries (OC). This result is expected taking into account the short periods of the binaries (Table~\ref{Tab1}).

(2) The $q$ values are near 0.5 for two binaries (targets 1 and 6) and near 1 for the rest ones. This result refers to our sample only and should be assumed as accidental because the $q$ values of W UMa stars are rather evenly distributed (fig. 1 in \citealt{Csizmadia+Klagyivik+2004}).


(3) The stellar components are of K spectral type. This is a natural consequence of the average infrared colors of the targets (Table~\ref{Tab4}).

(4) The temperature differences of the components of the binaries are below 630 K. The small temperature differences are expected for contact or overcontact systems and means that their components are almost in thermal contact.

(5) Four targets (2, 3, 4, 5) have equal in size components. Just they have mass ratio near 1. Hence, their components obey the same mass-radius relation.

\newpage
(6) Five targets revealed O'Connell effect that was reproduced by cool spots (Table~\ref{Tab6}) on the side surfaces of their primary components. The light curves of target 2 in 2011 and 2012 were fitted by diametrically opposite spots (visible at the second and first quadratures correspondingly). This result can be explained by differential rotation or by new spot cycle. But the equal sizes and temperatures of the two spots (Table~\ref{Tab6}) support rather the first supposition.

The modeling of the O'Connell effect of late stars by cool spots is preferred approach of almost all authors. Cloud of circumstellar, absorbing matter that orbits the corresponding star may reproduce the O'Connell effect as the cool photospheric spot. Only photometric data (as the ordinary case is) cannot resolve this ambiguity and one adopts modeling by spots.

(7) The cool spot of target 5 was invoked to reproduce the distortions of its light curve at the minima and maxima.

\begin{figure}
   \centering
   \includegraphics[width=7.5cm, angle=0]{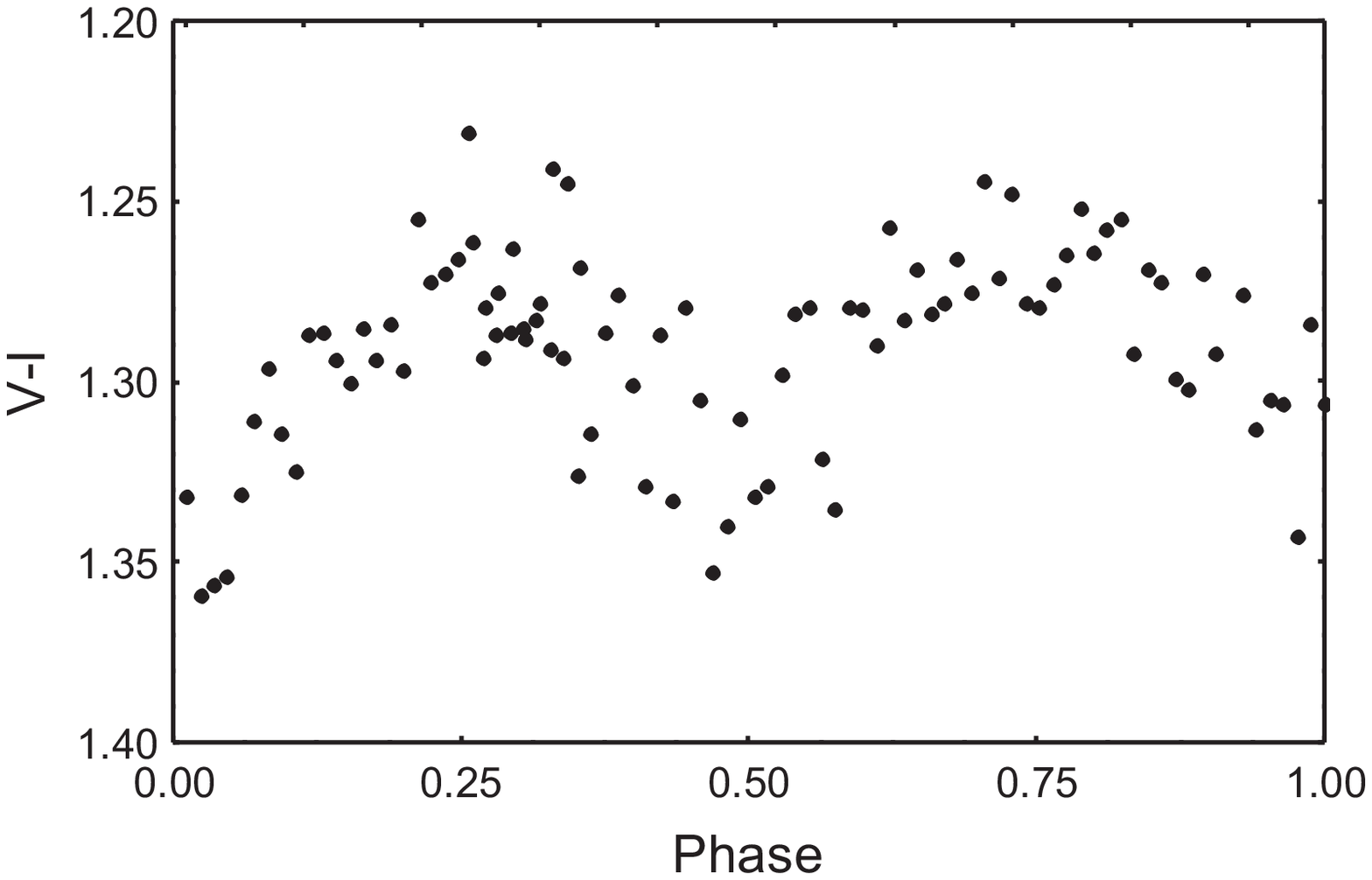}
   \caption{$V-I$ light curve of target 1}
   \label{Fig9}
   \end{figure}

\begin{table}
\bc
\begin{minipage}[]{100mm}
\caption[]{Amplitudes of the reddened $(V-I)$ colors and amplitudes of $I$ variability of the targets\label{Tab7}}\end{minipage}
\setlength{\tabcolsep}{1pt}
\small
 \begin{tabular}{ccccccc}
  \hline\noalign{\smallskip}
Target        & 1    & 2    & 3    & 4    & 5    &  6  \\
  \hline\noalign{\smallskip}
$\Delta (V-I)$& 0.06 & 0.08 & 0.10 & 0.14 & 0.03 & 0.07\\
$\Delta I$    & 0.30 & 0.55 & 0.55 & 0.55 & 0.21 & 0.32\\
  \noalign{\smallskip}\hline
\end{tabular}
\ec
\end{table}

\newpage
(8) The \emph{V-I} indices of the targets undergo orbital variability with reddening at the light minima (Fig.~\ref{Fig9}). The amplitudes of the \emph{V-I} curves are proportional to the light amplitudes (Table~\ref{Tab7}). This result means that the temperature of the back surface of the close stellar components is
lower than those of their side surfaces.

(9) No target of our sample was identified as an X-ray source and H$_{\alpha}$ emission source. Thus the low activity of the W UMa stars was confirmed.

\section{Subclassification of our targets}

The contact binary stars were divided into two sub-types, A and W, according to the following criteria: (a) the ratio \emph{R/T} (radius to temperature): the larger star is the hotter one for A-type system while the smaller star is the hotter one for W-type system (\citealt{Binnendijk+1970}); (b) temperature \emph{T} or spectral type: the A type systems are earlier than the W type binaries which components are of G and K spectral type; (c) period \emph{P}: the W-type binaries have shorter periods of 0.22 to 0.4 day (\citealt{Smith+1984}); (d) mass ratio \emph{q}: values of A type are smaller than those of W type.




\newpage
\cite{Csizmadia+Klagyivik+2004} introduced H subtype systems (H/A and H/W) with large mass ratio ($q\geq$ 0.72) which energy transfer is less efficient than that in other types of contact binary stars. They found that the different subtypes of W UMa's are located into different regions on the mass ratio -- luminosity ratio diagram but above the line $\lambda = q^{4.6}$ ($\lambda = l_2/l_1$) representing the mass-luminosity relation for the MS stars (fig. 1 in \citealt{Csizmadia+Klagyivik+2004}).

The classification of our targets according to the foregoing criteria is presented in Table~\ref{Tab8}. It reveals that almost all targets belong simultaneously at least to the two subclasses. Only target 3 is deeply below the line representing the mass-luminosity relation for the MS stars (uninhabited place)
This result means that the proposed criteria and subdivision are ambiguous and, of course, that the stellar world is richer than one expects.

\begin{table}
\bc
\begin{minipage}[]{100mm}
\caption[]{Subclassification of the targets\label{Tab8}}\end{minipage}
\setlength{\tabcolsep}{1pt}
\small
 \begin{tabular}{ccccccc}
  \hline\noalign{\smallskip}
Criterion/target    & 1 & 2 & 3 & 4 & 5 & 6   \\
  \hline\noalign{\smallskip}
$T/R$            & A & A   & A & A   & A   & A \\
$T$              & W & W   & W & W   & W   & W \\
$P$              & W & W   & W & W   & W   & W \\
$q$              & A & W   & W & W   & W   & A \\
$q, l_{2}/l_{1}$ & A & H/A & ? & H/A & H/W & A \\
  \noalign{\smallskip}\hline
\end{tabular}
\ec
\end{table}

\newpage
\section{Conclusion}

We obtained light curve solutions of six binaries with periods around 0.25 d, the lower limit of the periods of the W UMa stars. Three of the targets were newly discovered eclipsing systems.

The solutions revealed that all systems are contact or overcontact binaries.

The colors of the investigated systems become redder at the light minima. Such a behavior may mean that the temperatures of the back surfaces of the close components are lower than those of their side surfaces.

The mass ratios of the targets are grouped around values 0.5 and 1 and those with mass ratio near 1 have equal in size components.

Five targets revealed O'Connell effect that was reproduced by cool spots on the side surfaces of their primary components. The light curves of V1067 Her in 2011 and 2012 were fitted by diametrically opposite spots.

The applying of the criteria for subdivision of the W UMa stars to our targets led to ambiguous results. New criteria are necessary for the subdivision of the numerous W UMa systems.

This study adds new six systems with estimated parameters to the family of short-period binaries. They could help to improve the statistical relations between the stellar parameters of the low-mass stars.

\normalem
\begin{acknowledgements}
The research was supported partly by funds of projects RD 02-263 of Scientific Foundation of Shumen University.

This publication makes use of data products from the Two Micron All Sky Survey, which is a joint project of the University of Massachusetts and the Infrared Processing and Analysis Center/California Institute of Technology, funded by the National Aeronautics and Space Administration and the National Science
Foundation. This research also has made use of the SIMBAD database, operated at CDS, Strasbourg, France, NASA's Astrophysics Data System Abstract Service, and the USNOFS Image and Catalogue Archive operated by the United States Naval Observatory, Flagstaff Station (http://www.nofs.navy.mil/data/fchpix/).

The authors are grateful to the anonymous reviewer for the useful suggestions and notes.
\end{acknowledgements}

\newpage
\bibliographystyle{raa}
\bibliography{bibtex}

\end{document}